# Enhanced Computation Method of Topological Smoothing On Shared Memory Parallel Machines

Ramzi MAHMOUDI[1], Mohamed AKIL[1]

[1]*IGM, Unité Mixte CNRS-UMLV-ESIEE UMR8049, University Paris-Est*
*Cité Descartes, BP99, 93162 Noisy Le Grand, France*
{mahmoudr, akilm}@esiee.fr

## Abstract

To prepare images for better segmentation, we need preprocessing applications, such as smoothing, to reduce noise. In this paper, we present an enhanced computation method for smoothing 2D object in binary case. Unlike existing approaches, proposed method provides a parallel computation and better memory management, while preserving the topology (number of connected components) of the original image by using homotopic transformations defined in the framework of digital topology. We introduce an adapted parallelization strategy called split, distribute and merge (SDM) strategy which allows efficient parallelization of a large class of topological operators. To achieve a good speedup and better memory allocation, we cared about task scheduling and managing. Distributed work during smoothing process is done by a variable number of threads. Tests on 2D grayscale image (512*512), using shared memory parallel machine (SMPM) with 8 CPU cores (2× Xeon E5405 running at frequency of 2 GHz), showed an enhancement of 5.2 with cache success rate of 70%.

## 1   Introduction

Smoothing filter is the method of choice for image preprocessing and pattern recognition. For example, the analysis or recognition of a shape is often perturbed by noise, thus the smoothing of object boundaries is a necessary preprocessing step. Also, when warping binary digital images, we obtain a crenellated result that must be smoothed for better visualization. The smoothing procedure can also be used to extract some shape characteristics: by making the difference between the original and the smoothed object, salient or carved parts can be detected and measured.

Smoothing shape has been extensively studied and many approaches have been proposed. The most popular one is the linear filtering by Laplacien smoothing for 2D-vector [1] and 3D mesh [2].  Other approach by morphological filtering can be applied directly to the shape [3] or to curvature plot of the object's contour [4]. Unfortunately none of these operators preserve the topology (number of connected components) of the original image. In 2004, our team introduced a new method for smoothing 2D and 3D objects in binary images while preserving topology [5]. Objects are defined as sets of grid points, and topology preservation is ensured by the exclusive use of homotopic transformations defined in the framework of digital topology [6]. Smoothness is obtained by the use of morphological openings and closings by metric discs or balls of increasing radius, in the manner of alternating sequential filters [7]. The authors' efforts have brought about two major issues such as preserving the topology and the multitude of objects in the scene to smooth out without worrying about memory management, latency or cadency of their filter. This paper describes an enhanced computation method of topological smoothing filter that assure better performance. We present also a new parallelization strategy, called Split Distribute and Merge (SD&M). Our strategy is designed specifically for topological operator's parallelization on shared memory architectures. The new strategy is based upon the exclusive combination of two patterns: divide and conquer and event-based coordination.

This paper is organized as follows: in section 2, some basic notions of topological operators are summarized; the original smoothing filter is introduced. In section 3, parallelization strategy, that has been adopted, is introduced. We define the class of operators that our strategy may cover. Motivations for using SMP machines are also cited. Threads coordination and tasks scheduling are discussed. In section 4, the new parallel smoothing method is introduced and evaluations of acceleration, efficiency and success rate of cache memory access are also presented and discussed. Finally, we conclude with summary and future work in section 5.

## 2    Theoretical background

In this section, we recall some basic notions of digital topology [6] and mathematical morphology for binary images [8]. We define also the homotopic alternating sequential filters [5]. For the sake of simplicity, we restrict ourselves to the minimal set of notions that will be useful for our purpose. We start by introducing morphological operators based on structuring elements which are balls in the sense of Euclidean distance, in order to obtain the desired smoothing effect.

We denote by $\mathbb{Z}$ the set of relative integers, and by $E$ the discrete plane $\mathbb{Z}^2$. A point $x \in E$ is defined by $(x_1, x_2)$ with $x_i \in \mathbb{Z}$. Let $x \in E, r \in \mathbb{N}$, we denote by $B_r(x)$ the ball of radius $r$ centred on $x$, defined by $B_r(x) = \{y \in E, d(x, y) \leq r\}$, where $d$ is a distance on $E$. We denote by $B_r$ the map which associates to each $x$ in $E$ the ball $B_r(x)$. The Euclidean distance $d$ on $E$ is defined by: $d(x, y) = \left[ A^2 - B^2 \right]^{1/2}$ with $A = (x_1 - y_1)$ and $B = (x_2 - y_2)$.

An operator on $E$ is a mapping from $P(E)$ into $P(E)$, where $P(E)$ denotes the set of all subsets of $E$. Let $r$ be an integer, the dilation by $B_r$ is the operator $\delta_r$ defined by $\delta_r(X) = \bigcup_{x \in X} B_r(x) \ \forall X \in P(E)$. The ball $B_r$ is termed as the structuring element of the dilation. The erosion by $B_r$ is the operator $\varepsilon_r$ defined by duality: $\varepsilon_r = *\delta_r$.

Now, we introduce notion of simple point which is fundamental for the definition of topological operators in discrete spaces. We give a definition of local characterization of simple points in $E = \mathbb{Z}^2$. Let consider two neighbourhoods relations $\Gamma_4$ and $\Gamma_8$ defined for each point $x \in E$ by:

$$\Gamma_4(x) = \{y \in E; |y_1 - x_1| + |y_2 - x_2| \leq 1\}, \ \Gamma_8(x) = \{y \in E; \max |y_1 - x_1|, |y_2 - x_2| \leq 1\}.$$

For general case, we define $\Gamma_n^*(x) = \Gamma_n(x) \setminus \{x\}$ with $n \in \{4, 8\}$. Thus $y$ is said n-adjacent to $x$ if $y \in \Gamma_n^*(x)$. We say also that two points $x$ and $y$ of $X$ are n-connected in $X$ if there is a n-path between these two points. The equivalence classes for this relation are n-connected components of $X$. A subset $X$ of $E$ is said to be n-connected if it consists of exactly one n-connected component. The set of all n-connected components of $X$ which are n-adjacent to a point $x$ is denoted by $C_n[x, X]$. In order to have a correspondence between the topology of $X$ and the topology of $\overline{X}$, we use n-adjacency for $X$ and $\overline{n}$-adjacency for $\overline{X}$, with $(n, \overline{n})$ equal to (8; 4) or (4; 8).

Informally, a simple point $p$ of a discrete object $X$ is a point which is inessential to its topology. In other words, we can remove $p$ from $X$ without changing its topology. A point $x \in X$ is said simple if each n-component of $X$ contains exactly one n-component of $X \setminus \{x\}$ and if each $\overline{n}$-component of $\overline{X} \cup \{x\}$ contains exactly one $\overline{n}$-component of $\overline{X}$. Let $X \subset E$ and $x \in E$, two connectivity numbers defined as follows (# $X$ = cardinality of $X$ ): $T(x, X) = \#C_n\left[x, \Gamma_8^*(x) \cap X\right]; \overline{T}(x, X) = \#C_{\overline{n}}\left[x, \Gamma_8^*(x) \cap \overline{X}\right]$.

The following properties allows us to locally characterize simple points [6][9] hence to implement efficiently topology preserving operators: $x \in E$ is simple for $X \subseteq E \leftrightarrow T(x, X) = 1$ and $\overline{T}(x, X) = 1$.

The homotopic alternating sequential filter is a composition of homotopic cuttings and fillings by balls of increasing radius. It takes an original image $X$ and a control image $C$ as input, and smoothes $X$ while respecting its topology and geometrical constraints implicitly represented by $C$. A simple illustration is given by figure 1. Smoothed image (b) is obtained using HAS filter with a radius equal to five and four connectedness ($\Gamma_4$). More example can be found in [5].

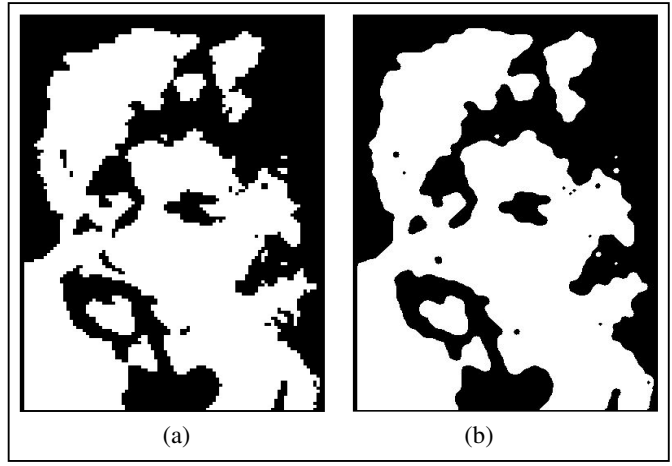

**Fig.1**: (a) Input image (b) Smoothed image

Based on this filter, Authors [5] introduce a general smoothing procedure with a single parameter to control smoothing degree. Let $C \subseteq X$, $r \in \mathbb{N}$ and $D \subseteq \overline{X}$ with $X$ any finite subset of $E$. The homotopic alternating sequential filter ($HASF$) of order $n$, with constraint sets $C$ and $D$, is defined as follows:

$$HASF_n^{C,D} = HF_n^D \circ HC_n^C \circ ... HF_1^D \circ HC_1^C$$

In the previous formula, $HC_n^C$ (i) refers to homotopic cutting of $X$ by $B_n$ with constraint set $C$ and $HF_n^D$ (ii) refers to homotopic filling of $X$ by $B_n$ with constraint set $D$. These two homotopic operators can be defined as follows:

$$HC_n^C(X) = *H(Y,V) \text{ With } \begin{cases} Y = H(X, \varepsilon_n(X) \cup C) \\ V = (\delta_n(Y) \cap X) \end{cases} \text{(i)} \qquad HF_n^D(X) = H(Z,W) \text{ With } \begin{cases} Z = *H(X, \delta_n(X) \cap \overline{D}) \\ W = (\varepsilon_n(Y) \cup X) \end{cases} \text{(ii)}$$

We recall that $H(Z,W)$ is an homotopic constrained thinning operator. It gives the ultimate skeleton of $Z$ constrained by $W$. The ultimate skeleton is obtained by selecting simple point in increasing order of their distance to the background thanks to a pre-computed Euclidian distance map [10]. We recall also that $*H(Y,V)$ is an homotopic constrained thickening operator. It thickens the set of $Y$ by iterative addition of points which are simple for $\overline{Y}$ and belong to the set $V$ until stability.

We have provided, in this section, the theoretical underpinnings for studying topological transforms. We introduced also the homotopic alternating sequential filter which constitutes base for topological smoothing.

## 3 Parallelization Strategy

In this section, we start by defining the class of topological operators. We also present our motivation to parallelize these algorithms on parallel shared memory machines. Then, we will introduce different steps of our approach after making a brief classification over existing strategies. We will focus especially on distribution phase and tasks scheduling over different processors. Scheduling and merging algorithms are presented and discussed. To illustrate both algorithms, scenarios are also introduced and discussed.

## 3.1 Class of topological algorithms

In 1996, Bertrand and Couprie [11] introduced connectivity numbers for grayscale image. These numbers describe locally (in a neighborhood of 3x3) the topology of a point. According to this description any point can be characterized following its topology. They also introduced some elementary operations able to modify gray level of a point without modifying image topology. These elementary operations of point characterization present a fundamental link between large class of topological operators including, mainly, skeletonization and crest restoring algorithms [12]. This class can also be extended, under condition, to homotopic kernel and leveling kernel transformation [13], topological watershed algorithm [14] and topological smoothing algorithm [5] which is the subject of this article. All mentioned algorithms get also many algorithmic structure similarities. In fact associated characterizations procedures evolve until stability which induce common recursion between different algorithms. The grey level of any point can also be lowered or enhanced more than once. Finally, all mentioned algorithms get a pixel's array as input and output data structure. It is important to mention that, to date, this class has not been efficiently parallelized like other classes as connected filter of morphological operator which recently has been parallelized in Wilkinson's work [15]. Parallelization strategy proposed by Seinstra [30] for local operators and point to point operators can also be cited as example. For global operators, Meijster strategy [16] shows also consistence. Hence the need of a common parallelization strategy for topological operators that offers an adapted algorithm structure design space. Chosen algorithm structure patterns that will be used in the design must be suitable for SMP machines.

In reality, although the cost of communication (Memory-processor and inter-processors) is high enough, shared memory architectures meet our needs for different reasons: (i) These architectures have the advantage of allowing immediate sharing of data with is very helpful in the conception of any parallelization strategy (ii) They are non-dedicated architecture using standard component (processor, memory ...) so economically reliable (iii) They also offer some flexibility of use in many application areas, particular image processing.

## 3.2 Split Distribute and Merge Strategy

In practice, most effective parallel algorithm design might make use of multiple algorithm structures thus proposed strategy is a combination of the divide and conquer pattern and event-based coordination pattern, see figure 2. Hence the name that we have assigned: SD&M (Split Distribute and Merge) strategy. Not to be confused with mixed-parallelism approach (combining data-parallelism and task-parallelism), it is important to mention that our strategy (i) represents the last stitch in the decomposition chain of algorithm design patterns and it provides a fine-grained description of topological operators parallelization while mixed-parallelism strategy provides a coarse-grained description without specifying target algorithm. (ii) It covers only the case of recursive algorithms, while mixed-parallelization strategy is effective only in the linear case. (iii) It is especially designed for shared memory architecture with uniform access.

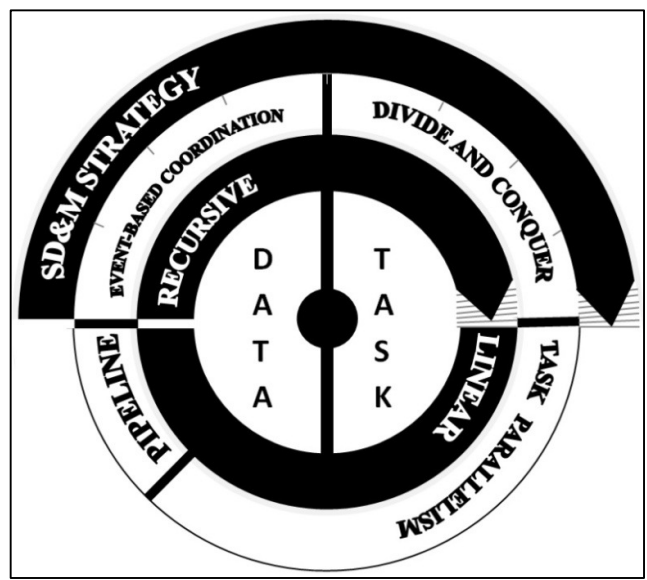

**Fig.2**: SDM strategy classification

### 3.2.1 Split phase

The Divide and Conquer pattern is applied first by recursively breaking down the problem into two or more sub-problems of the same type, until these become simple enough to be solved directly. Splitting the original problem take into account, in addition to the original algorithm's characteristics (mainly topology preservation), the mechanisms by which data are generated, stored, transmitted over networks (processor-processor or memory-processor), and passed between different stages of computation.

### 3.2.2 Distribute phase

Work distribution is a fundamental step to assure a perfect exploitation of multi-cores architecture's potential. We'll start by recalling briefly some basic notion of distribution techniques then we introduce our minimal distribution approach that is particularly suitable for topological recursive algorithms where simple point characterization is necessary. Our approach is general and applicable to shared memory parallel machines. Critical cases are also introduced and discussed.

Indeed there are two main types of scheduler. There are those designed for real-time systems (RTS). In this case, the most commonly approaches used to schedule real-time task system are: Clock-Driven, Processor-Sharing and Priority-Driven. Further description of different scheduling approaches can be found in [26,27,28]. According to [28] the Priority-Driven is far superior the other approaches. These schedulers must provide an operational RTS: completed work and delivered results on a timely basis. Other schedulers are designed for Non Real-time system. In this case, schedulers are not subject to the same constraints. Thus, "Symmetric Multiprocessing" scheduler distributes tasks to minimize total execution time without load balancing between processors, see figure 3 (a). On multi-core architectures, this can lead to high occupancy rate of one processor while the others are free.

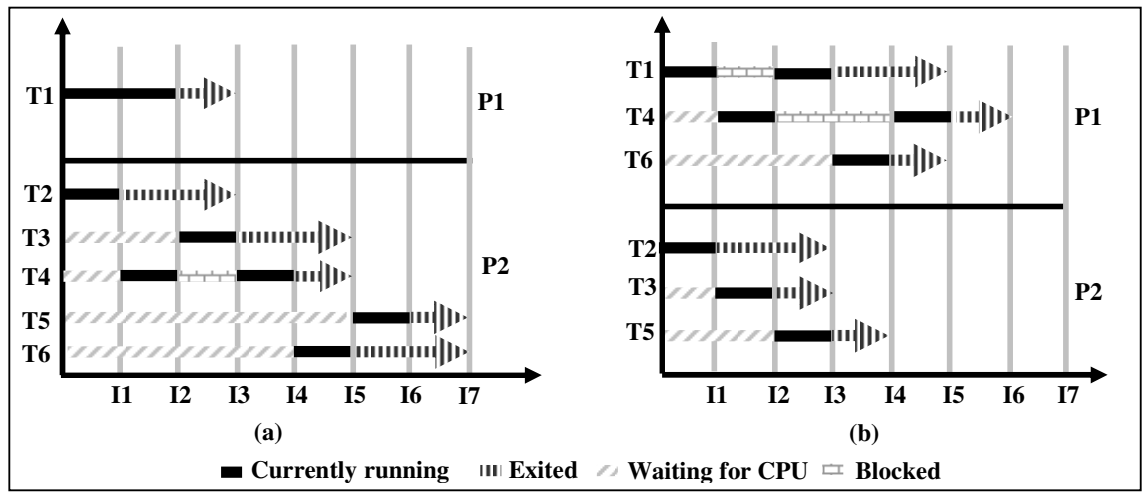

**Fig.3**: (a) Non-real time symmetric task distribution (b) Task distribution based upon uniformity principle

We propose a novel tasks scheduling approach to prevent improper load distribution while improving total execution time, see figure 3 (b). In literature, there are several schedulers that provide a balanced distribution of tasks such as RSDL "Rotating Staircase Deadline" [17] which incorporates a foreground-background descending priority system (the staircase) with run-queue managing minor and major epochs (rotation and deadline). Other scheduler, as CFS "Completely Fair Scheduler" [18], shows consistence. It handles resource allocation for executing processes, and aims to maximize overall CPU utilization while maximizing interactive performance. These schedulers are based on tasks uniformity principle. Through the tasks homogeneity, better distribution can be achieved and total execution time reduced. Unfortunately, these schedulers are not available in all operating system versions especially for small system. Based on the same principle of tasks uniformity, we propose a new scheduling algorithm, simpler to implement and more adapted to topological algorithm implementation.

Let be a basic non-preemptive scheduler 'Basic-NPS', $T = \{t_1, t_2, ..., t_k\}$ is the set of all tasks, $T_T = \{t_1, t_2, ..., t_i\}$ is the set of tasks to process with $T_T \subset T$, $P = \{p_1, p_2, ..., p_n\}$ is the set of all processors and $P_a = \{p_1, p_2, ..., p_j\}$ is the set of available processors with $P_a \subset P$.

Basic-NPS ($T_x \Rightarrow P_y$) is able to schedule a set of $T_x$ tasks on $P_y$ processor. Let $\{p\}$ be the maximum of processors that $P_y$ will contain. Then $\{p\}$ can be defined as the maximum of available processors already defined by the set $P_a$ and $\{p\} = \max p_j / p_j \in P_a$. While $([P_a \neq \emptyset] \wedge [T_T \neq \emptyset])$ then $T_x \Rightarrow P_y$ : $T_x \in T_T$; $P_y \in P_a$.

In this scheduler, each processor will treat at maximum $m = \max t_i / t_i \rightarrow p_j \leq \max(\frac{|T|}{|P|})$ tasks with $j \in \{1, 2, ..., n\}$. Then, the worst case to process $T$ is $K(T) = \max \left\{ \max_i T_T \rightarrow p_1, ..., \max_{j<...<i} T_T \rightarrow p_k \right\}$. As proof, let suppose that it exist a set $L(T)$ as $\sum L(T) \geq \sum K(T)$. As 'Basic-NPS' manage $L(T)$ and $K(T)$, so we can introduce the following: $|L(T)| \leq m$ and $|K(T)| \leq m$. Thus, if $(\sum L(T) \geq \sum K(T))$ then it exists at least one task $\{l\}$, with $k \in K(T)$, such as: $(A \wedge B \wedge C)$ with $A = (l \in L(T)), B = (l \notin K(T)), C = (l > k)$. This is impossible according to the definition of $K(T)$ which was defined as the worst case.

---

**Algorithm 1: Scheduling policy**

1. $T$ : Set of all tasks
2. $P$ : Set of all processors
3. While $(T \neq \emptyset)$ repeat :
4.     $N_T$ = Nbr_active_tasks() ;
5.     $N_P$ = Nbr_ available_processors();
6.     If $(N_P \neq 0)$ then
7.       If ( $N_T < N_P$ ) then
8.          For each processor $N_{Pi}$ :
9.            Generate-new-process ( $N_{Ti}$ );
10.            Identify-class ( $N_{Ti}$ , SCHED_FIFO);
11.          Endfor
12.       Else : $N_{DT}$ = Desable_tasks ( $N_P - N_T$ ) ;
13.          Insert_desabled_tasks ( $N_{DT}$ ,$T$ );
14.          For each processor $N_{Pi}$ :
15.            Generate-new-process ( $N_{Ti}$ );
16.            Identify-class ( $N_{Ti}$ , SCHED_FIFO);
17.          Endfor
18.      EndIf
19.    EndIf
20. EndWhile

Algorithm 1 describes 'Basic-NPS' policy. The first step consists on asking operating system to determine the number of available processor. Depending on this number, algorithm will generate process. One active process will be assigned for each available processor. These new processes will belong to the SHED_FIFO class in order to ensure preemption and especially to avoid context switching. Process will only stop running if work is complete or less frequently when another process, belonging to the same class, with higher priority requesting processor. The global execution will stop if there no more task to process.

### 3.2.3 Merging phase

The key problem of each parallelization is merging obtained results. Normally this phase is done at the end of the process when all results are returned by all threads what usually means that only one output variable is declared and shared between all fighting threads. But as we mentioned in section 3.1, we are dealing with a dynamic evolution and if we take into account different steps of simple point detection then pixel characterizations, we can plan the following: The original shared data structure, containing all pixels, is divided into $n$ research zones $\{z_1, z_2..., z_n\}$. We associate one thread from the following list $\{T_1, T_2..., T_n\}$ to each zone. Each thread can browse freely its zone and if it detects target pixel types, it lowers characterized pixel and it pushes its eight neighbors in one of the available FIFO queues. A queue is said available if only one thread (owner) is using it. One queue cannot be shared by more than two threads so if no queue is available, threads can create a new one and become owners.

Since two threads finished, they directly merge and a new thread is created and then same process is lunched again. New created thread will inherit queue shared between his parents. Thus it can restart research. It is also important to mention that there is no hierarchical order in thread merging, only criteria is finishing time. We mention also that one neighbor cannot be inserted twice. It is a precaution in order to minimize consumed cache. More formal description of merging techniques is given in by algorithm 2.

It is important to highlight similarity and difference that may exist between our merging algorithm and KPN [29]. In effect, both are deterministic and do not depend on execution order. But KPN algorithm may be executed in sequentially or in parallel with the same outcome while our merging algorithm is designed only for parallel execution. KPN support recurrence and recursion while our merging algorithm support only recursion.

In large scale application, KPN showed consistence. Examples include Daedalus project [31] where generated KPN models are used to map process into FPGA architecture. Ambric architectures [32] implement also a KPN model using bounded buffers to create massively DMP Machines based on structural object programming model.

In a narrower framework limited to simple point characterization, the implementation of such a model will be very expensive and it would be better to find an easier and more specific algorithm.

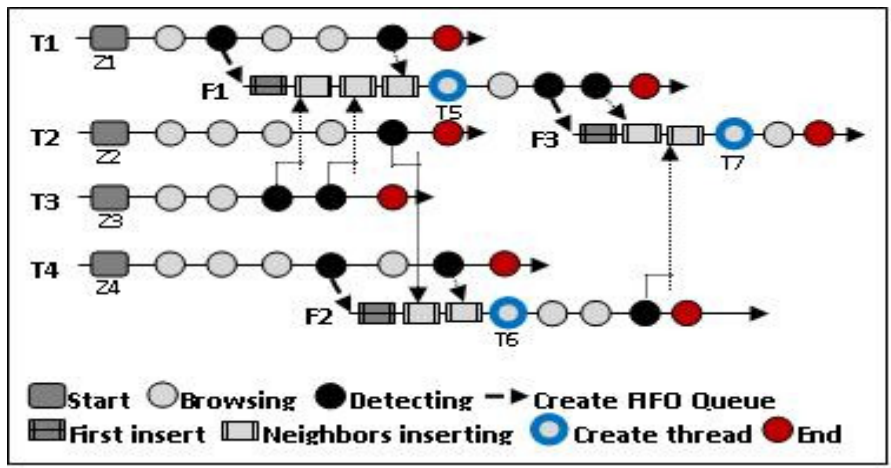

**Fig.4:** Illustration of merging phase with four threads

In Figure 4, we give an illustration of the merging algorithm with four threads. The original shared data structure is divided into 4 research areas $\{z_1, z_2, z_3, z_4\}$. Threads $\{T_1, T_2, T_3, T_4\}$ will start browsing different zones in parallel. $T_1$ is the first to detect target point (constructible, destructible …) so it lowers characterized pixel (in $z_1$) and it pushes its eight (or four) neighbors in FIFO queue $F_1$ that it has created before continue browsing. Later, $T_3$ will detect new target point so it will lower characterized pixel (in $z_3$) then push neighbors in $F_1$ before continue browsing. $T_3$ don't need to create new FIFO queue since $F_1$ is available. $T_1$ and $T_3$ will repeat this procedure twice. Since they finish browsing, they merge and new thread $T_5$ is born. $T_5$ will start browsing only $F_1$. Since it detect new target point so it will lower characterized pixel (in $z_5 = z_1 + z_3$) then push neighbors in $F_3$ that it has created before continue browsing. Similarly $T_2$ and $T_4$ will generate the creation of $F_2$ and $T_6$. Here $T_6$ will eventually merge with $T_5$ to give birth to $T_7$. Finally there will be a single thread $T_7$ which will brows $F_3$ without detection any target points.

| **Algorithm 2: Merging technique** |
|---|
| 1.    $Z$ : Set of research zones |
| 2.    $T$ : Set of threads |
| 3.    $FIFO\_Q$ : Set of available FIFO queues |
| 4.    $P_T$ : Target pixel type ; $P_D$ : Detected pixel |
| 5.    For all zones $(Z_i \in Z)$ do: |
| 6.      Parallel_browsing ($T_i, Z_i$) ; |
| 7.    EndFor |
| 8.    For each thread $(T_i \in T)$ do: |
| 9.      If (pixel_caract($T_i, P_T$)==True) then |
| 10.        modify_value($P_D$); |
| 11.      If ($(FIFO\_Q \neq \varnothing)$ then |
| 12.        usedstatus($FIFO\_Q_j$,true) ; |
| 13.        insert_neighbors($T_i, P_D, FIFO\_Q_j$); |
| 14.      Else : add_new_fifo ($FIFO\_Q$) |
| 15.        usedstatus($FIFO\_Q_{j+1}$,False) ; |
| 16.        insert_neighbors($T_i, P_D, FIFO\_Q_{j+1}$); |
| 17.      EndIf ; |
| 18.    EndIf ; |
| 19. EndFor; |

We have introduced, in this section, three necessary steps to implement our parallelization strategy (SDM). It is important to mention that some similarity may exist between our split/merging phases and alpha-extension/beta-reduction phases from structural perspective. Actually both approaches intended to put in place more guarantees that the parallelism will actually be met. But uses contexts are different. In effect, Jean Paul Sansonnet [33,34,35] team introduced alpha-extension (diffusion) and beta-reduction (merging) notions for stream manipulation in the framework of Declarative Data Parallel language definition and there techniques cannot be applied without a scalar function. While our proposal is restricted to topological characterization in the framework of topological operator's parallelization and no scalar function is required during the application of these two phases.

## 4 Parallel smoothing filter

In this section we start by analyzing overall structure of original algorithm. Then we continue with the parallelization of Euclidean distance, thinning and thickening algorithm. We conclude by a performance analysis of the entire smoothing topological operator. Obtained execution time, efficiency, speedup and cache misses will be introduced and discussed.

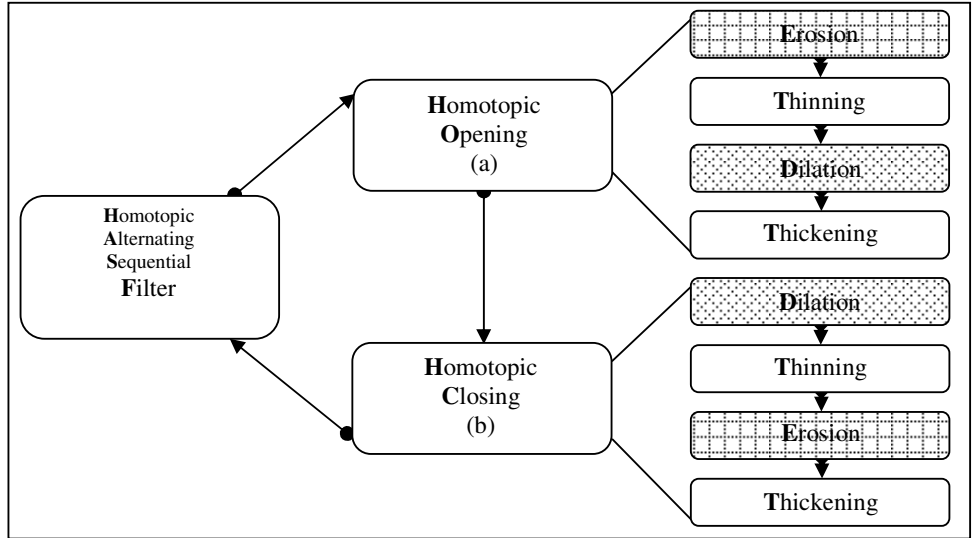
**Fig.5**: Overall structure of the original smoothing algorithm

As we have shown in Section 2, smoothing algorithm receives as input a binary image and maximum radius. It uses two procedures for homotopic opening and closing, see figure 5 (a) (b). The call is looped to ensure an ongoing relationship between input and output. The opening process is a consecutive execution of erosion, thinning, dilatation and thickening. While closure procedure ensures the same performance of the four consecutive functions with single difference: the erosion instead of dilatation. Thinning and thickening ensure the topological control of erosion and dilatation. This control is based on researching and removing of all destructible points. When destructible point is deleted, its neighbors are reviewed to ensure that they are not destructible either.

|  | 200x200 | | | 168x288 | | |
|---|---|---|---|---|---|---|
|  | r=5 | r=10 | r=∞ | r=5 | r=10 | r=∞ |
| **EucDis** (%) | 64.44 | 54.93 | 46.67 | 59.25 | 49.79 | 35.25 |
| **TopCar** (%) | 8.89 | 13.89 | 18.15 | 11.58 | 16.50 | 24.03 |

**Tab.1:** Time execution rate of E.D and topological characterization functions

A preliminary assessment of first implementation code, see Table 1, shows that Euclidean distance computing (**EucDis**) takes more time than topological point characterization (**Topcar**). For an image of (200*200), computation time of E.D with an infinite radius is 46.67% while point characterization of 2.4 million points occupies only 18.15%. If we limit radius between 5 and 10, computation time of E.D. continues to increase. It can reach 64.44% of total time with a radius equal to 5. However time for topological characterization is only 8.89% for 1 million points. These finding remain the same if we increase image size. Beyond (512*512), computing time of point characterization becomes considerable.

### 4.1 Euclidean distance computing

#### 4.1.1 Study on Euclidian Distance algorithms

During previous evaluation, 4SED [10] algorithm was used for Euclidean distance computation. So we are looking for another algorithm that is faster, and parallelizable. New algorithm must have an Euclidean distance computation error less than, or equal to, that produced by 4SED in order to maintain homotopic characteristics of the image. In literature, several algorithms for Euclidean distance computing exist. Lemire [19] and Shih [20] algorithms are bad candidates because Lemire's algorithm does not use Euclidean circle as structuring element. Then homotopic property will not be preserved. Shih's algorithm has a strong data dependency which penalizes parallelization. In [21], Cuissenaire propose a first algorithm for Euclidian distance computing, called PSN "Propagation Using a Single Neighborhood" that uses the following element structure:

$$d_4(p) = \left\{ q \vee \sqrt{(q_x - p_x)^2 + (q_y - p_x)^2} < 1 \right\} \quad \text{(i)}$$

He also proposes a second algorithm, called PMN "Propagation Using Multiple Neighborhood" that uses eight neighbors. In [22], he also proposes a third algorithm with $o(n^{3/2})$ complexity, which offers an accurate computation of the Euclidean distance. Only drawback of this third algorithm is computation time which is very important and goes beyond the two algorithms mentioned above. Even if computing error produced by PSN is greater than computing error produced by PMN, it is comparable to that produced by 4SED. Low data dependence and ability to operate on 3D images, makes PSN algorithm a potential candidate to replace 4SED.

Meijster [23] proposes an algorithm to compute exact Euclidean distance. Algorithm complexity is $o(n)$ and it operates in two independent, but successive, steps. First step is based on looking over columns then computing distance between each point and existing objects. Second step includes same treatment looking over lines. It is important to note that strong independence between different processing steps and computing error equal to zero makes Meijster algorithm another potential candidate to replace 4SED. Algorithm is also able to operate on 3D images. Theory analysis of Meijster and Cuissenaire algorithms can be found in Fabbri's work [24].

In the following, we propose first analysis based on different algorithms implementation in order to compare between them. We have implemented 4SED algorithm using a fixed size stack. This stack uses a FIFO queue and it has small size while 4SED algorithm does not need to store temporal image. Results are directly stored into the output image, we will retain this implementation because 4SED assessment serve only as reference for comparison. For PSN implementation, we used stacks with dynamic sizes. Memory is allocated using small blocks defined at stack creation. When an object is added to queue, algorithm will use available memory of last block. If no space is available, a new block is allocated automatically. Block size is proportional to image size (N x M / 100). Finally we used a simple memory structure to implement Meijster algorithm. A simple matrix was used to compute distance between points and object of each column and three vectors were used to compute distance in each line. We recall that this comparison is done in order to select the best algorithm among three candidates.

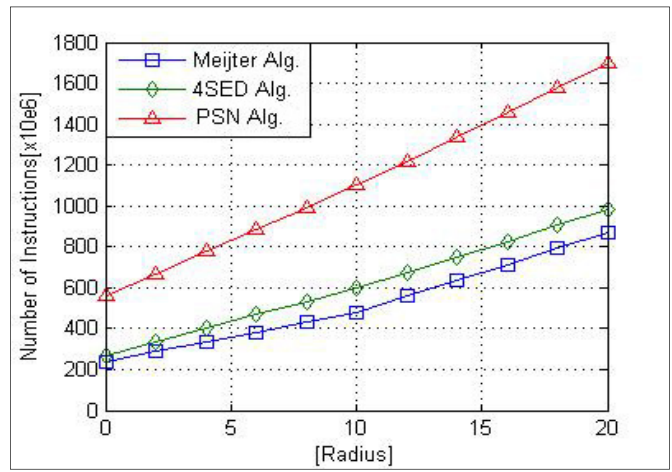

**Fig.6:** Comparison between different EDT algorithms

Figure 6 describes obtained results by different implementations on single processor architecture P4. During this evaluation we used binary test image (200x200). We have also varied ball radius. We used Valgrind software to evaluate different designs. Callgrind tool returns the cost of implementing of each program by detecting IF (Instruction Fetch). Results show that PSN algorithm is the most expensive in all cases (for any radius). Meijster algorithm is moderately faster than 4SED. The output images returned by Meijster algorithm hold the best visual quality while Euclidean distance computation error is almost zero thus our efforts will be brought on Meijster algorithm parallelization.

### 4.1.2 Parallelization of Meijster algorithm

We denote by $I$ input image with $m$ columns and $n$ rows. We denote by $B$ an object included in $I$. The idea is to compute, for each point $p \in I \wedge p \notin B$, separating distance between $p$ and the closest point $b$ with $b \in B$ and $\forall (0 \leq b \leq m)$, $b = (b_x, b_y)$. This amount to compute the following matrix: $dt[p_x, p_y] = \sqrt{EDT(p)}$ with $EDT(p) = \min(p_y - b_x)^2 + G(p_x, b_y)^2$.

If we assume that minimum distance of an empty group $K$ is $\infty$ and $\forall z \in K$, we have $(z_y + \infty) = \infty$ then $EDT(p)$ formula can be written as follow: $\forall b_x < n$, $\forall b_y \leq m$, $EDT(p) = \min(p_y - b_x)^2 + G(p_x, b_y)^2$ with $G(p_x, y) = \min |p_x - b_x| : b = (b_x, y)$.

Thus we can split the Euclidian distance transform procedure into two steps. The first step is to scan columns and compute $EDT$ for each column $y$. Second step consists on repeating the same procedure for each line. In the following we start by detailing these two steps: In the first step $G(p_x, y)$ can be computed through the two following sub functions: $G_T(p_x, y) = \min p_x - b_x : b = (b_x, y)$, $G_B(p_x, y) = \min b_x - p_x : b = (b_x, y)$ with $\forall 0 \leq b_x \leq n$. To compute $G_T(p_x, y)$ and $G_B(p_x, y)$, we scan each column $y$ from top to bottom using the two following formula: $G_T(p_x, y) = G_T(y, p_x - 1) + 1$  $G_B(p_x, y) = G_B(y, p_x + 1) + 1$. Thus sequential algorithm of the first step can be written as follows. Complexity order is $o(n \times m)$.

---

**Algorithm 3: Meijster O.V [1st Step]**

*Data* : m:colums, n:lines, b:image
1. **Forall** $y \in [0..m-1]$ **do**
2.     **If** $(0, y) \in B$ **then** $g[0..y] = 0$
3.         **else** $g[0..y] = \infty$
4.     **endif**

5.     */* $G_T$ */*
6.     **for** $(x = 1)$ **to** $(n-1)$ **do**
7.         **if** $[x, y] \in B$ **then** $g[x..y] = 0$
8.         **else** $g[x, y] = g[x+1, y] + 1$
9.         **endif**
10.    **endfor**

11.    */* $G_B$ */*
12.    **for** $(x = n-2)$ **downto** $(0)$ **do**
13.        **if** $g[x+1, y] < g[x, y]$ **then**
14.           $g[x, y] = g[x+1, y] + 1$
15.        **endif**
16.    **endfor**
17. **endforall**

---

Let's move to the second step. We start by defining $f(p, y) = (p_y - y)^2 + G(p_x, y)^2$. Then we can define $EDT(p) = \min f(p - y), \forall 0 \leq y \leq m$. For each row $u$, we note that there is, for the same point $p$, the same value of $f(p, y)$ for different values of $y$, so we can introduce the concept of "region of column ".

Let $S$ be the set of $y$ points such that $f(p,y)$ is minimal and unique. The formula of $S$, $\forall 0 \leq y \leq u$, is $S_p(u) = \min y : f(p,y) \leq f(p,i)$. $\forall 0 \leq i \leq u \wedge u \leq m$. Let $T$ be the set of points with coordinate greater than, or equal to, horizontal coordinate of the intersection with a region:
$T_p(u) = Sep_{p_x}(S_p(u-1), u) + 1$.

Let $Sep(i,u)$ be the separation between regions of $i$ and $u$, defined by:
$f(p,i) \leq f(p,u) \Leftrightarrow (p_y - i)^2 + G(p_x,i)^2 \leq (p_y - u)^2 + G(p_x,u)^2$
$\Leftrightarrow Sep_{p_x}(i,u) = (u^2 - i^2 + Dif)/2(u-i) = p_y$ With $Dif = (G(p_x,u)^2 - G(p_x,i)^2)$.

Thus lines will be processed, from left to right then from right to left. During the first term, from left to right, two vectors $S$ and $T$ will be created. These two vectors will contain respectively all regions and all intersections. During the second treatment, from right to left, we compute $f$ for each value of $S$. $f$ is also computed for each respective values of $T$. Algorithm 4 is associated to second step. For the first term, complexity order is $q + 2(m-u)$ whereas complexity order of the second term is only $m$.

The independence of data processing between rows and columns is the key to apply of SDM parallelization strategy. In the first stage, column processing, we can define data interdependence by the following equation:

$G(p_x, y) = \min\{G_T(p_x, y), G_B(p_x, y)\}$
$\Leftrightarrow G_T(p_x, y) = \begin{Bmatrix} 0 & if\ (p_x,y) \in B \\ G_T(p_x,y) & else \end{Bmatrix} \Leftrightarrow G_B(p_x, y) = \min\{G_B(p_x+1, y), G_T(p_x, y)\}$

It follows that values of each column y of G, depends only on lines: $p_x$, $p_x + 1$ and $p_x - 1$. Similarly, at the second stage, we can introduce the following interrelationship: $Edt(p) = f(p, S_p(q))$. Then $\forall (0 \leq y \leq u), (0 \leq i \leq u) \wedge (u < m)$, $S_p(u) = \min y : f(p,y) \leq f(p,i)$. Thus, if $(u = T_p(q))$ so $q = (q-1)$ which imply the following: $T_p(u) = Sep_{p_x}(S_p(q), u) + 1$.

| Algorithm 4: Meijster O.V [2nd Step] |
|---|
| *Data* : b:image, g: G_Table, m: columns, n:lines |
| *Result* : |
| 1.    Forall $x \in [0..n-1]$ do |
| 2.      $q = 0$ |
| 3.      $s[0] = 0$ |
| 4.      $t[0] = 0$ |
| 5.      /* *First part* */ |
| 6.      for $(u=1)$ to $(m-1)$ do |
| 7.        $A = (q \geq 0) \wedge \left[ f((x,t[q]), s[q]) \right]$ |
| 8.        $B = f((x,t[q]), u)$ |
| 9.        while $(A > B)$ then $q \leftarrow (q+1)$ |
| 10.        end while |
| 11. |
| 12.        if $(q < 0)$ then $(q \leftarrow 0)$ |
| 13.                 $(s[0] \leftarrow u)$ |
| 14.        else $w \leftarrow Sep(s[q], u, x) + 1$ |
| 15.          if $(w < m)$ then $q \leftarrow (q+1)$ |
| 16.             $s[q] \leftarrow u$ |

```
17.                   t[q] ← w
18.              endif
19.          endif
20.      endfor

21.      /* Second part */
22.      for (u = m−1) to (0) do
23.          Edt[x,u] = f((x,u), s[q])
24.          if (u = t[q]) then q ← (q−1)
25.          endif
26.      endfor
27. end forall
```

According to this formalization, values of $f(p,i)$ and $Sep_x(i,u)$ are independent of modified data. So using two vectors $S$ and $T$, a private variable $q$ for each line ensures complete independence in writing. We start applying the splitting step by sharing the columns and lines processing between multiple processors. A thread can process one or more columns and the number of threads used will depend on the number of processors. The results returned by all threads in this first stage will be merged in order to start lines processing. In the following we introduce the parallel version of Meisjter algorithm for both steps. Associated algorithm complexity is $o((n \times m)/N)$. $(n \times m)$ refers to image size and $N$ refers to the number of processors.

**Algorithm 5: Meijster P.V [1st step]**
```
1.  For (y = t, y < m, y = y + t_max) do
2.      If (0, y) ∈ B then g[0, y] ← 0
3.          else g[0, y] ← ∞
4.      endif
5.      /* G_T */
6.      for (x = 1) to (n−1) do
7.          if [x, y] ∈ B then g[x, y] ← 0
8.          else g[x, y] ← g[x+1, y] + 1
9.          endif
10.     Endfor

11.     /* G_B */
12.     for (x = n−2) downto (0) do
13.         if (g[x+1, y] < g[x, y]) then
14.             g[x, y] ← g[x+1, y] + 1
15.         endif
16.     endfor
17. Endforall
```

Proposed parallel version of Meijster algorithm was implemented in C using OpenMP directives. Speedup for numbers of threads equal to 1, 2, 4, 8, and 16 were determined. The efficiency measure $\Psi(n)$ is given by the following formula with $n$ the number of processors: $\Psi(n)$ = seq. time /($n$ *para. time) (ii)

Times were performed on eight-core (2× Xeon E5405) shared memory parallel computer, on Intel Quad-core Xeon E5335, on Intel Core 2 Duo E8400 and Intel mono-processor Pentium 4 660. The minimum value of 5 timings was taken as most indicative of algorithm speed. More information about architectures characteristics are given in Section 4.

**Algorithm 6: Meijster P.V [2nd Step]**

1.  For $(x = t, x < n, x = x + t_{max})$ do
2.     $q = 0$; $s[0] = 0$;
3.     $t[0] = 0$;
4.     /* First part */
5.     for $(u = 1)$ to $(m - 1)$ do
6.        $A \leftarrow (q \geq 0) \wedge [f((x, t[q]), s[q])]$
7.        $B \leftarrow f((x, t[q]), u)$
8.        while $(A > B)$ do $q \leftarrow (q + 1)$
9.        end while
10.
11.       if $(q < 0)$ then $(q \leftarrow 0)$
12.             $(s[0] \leftarrow u)$
13.       else $w \leftarrow Sep(s[q], u, x) + 1$
14.          if $(w < m)$ then $q \leftarrow (q + 1)$
15.             $s[q] \leftarrow u$
16.             $t[q] \leftarrow w$
17.          endif
18.       endif
19.    Endfor
20.    /* Second part */
21.    for $(u = m - 1)$ downto $(0)$ do
22.       $Edt[x, u] \leftarrow f((x, u), s[q])$
23.       if $(u = t[q])$ then $q \leftarrow (q - 1)$
24.       endif
25.    endfor
26. end forall

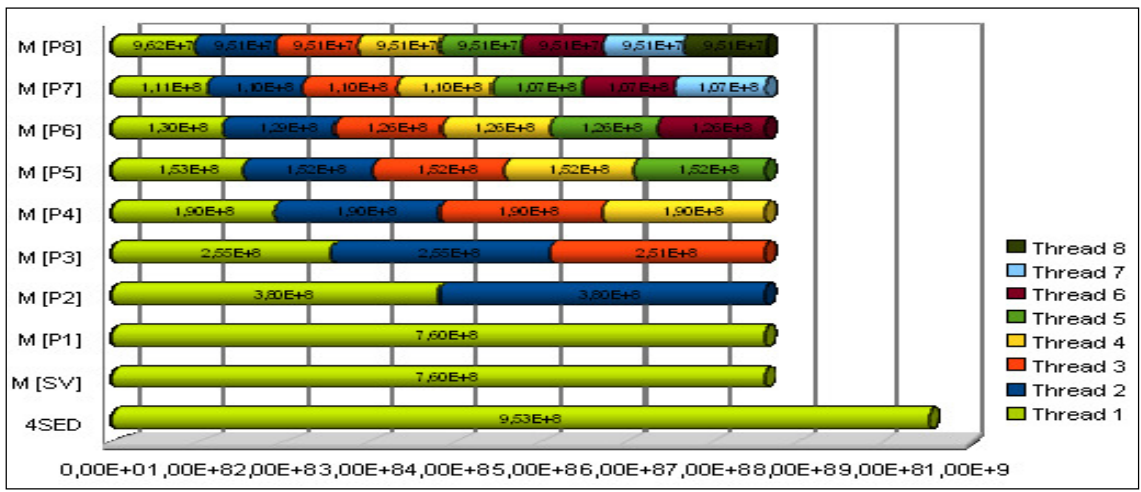

**Fig.7:** Instruction distribution (Meijster Alg)

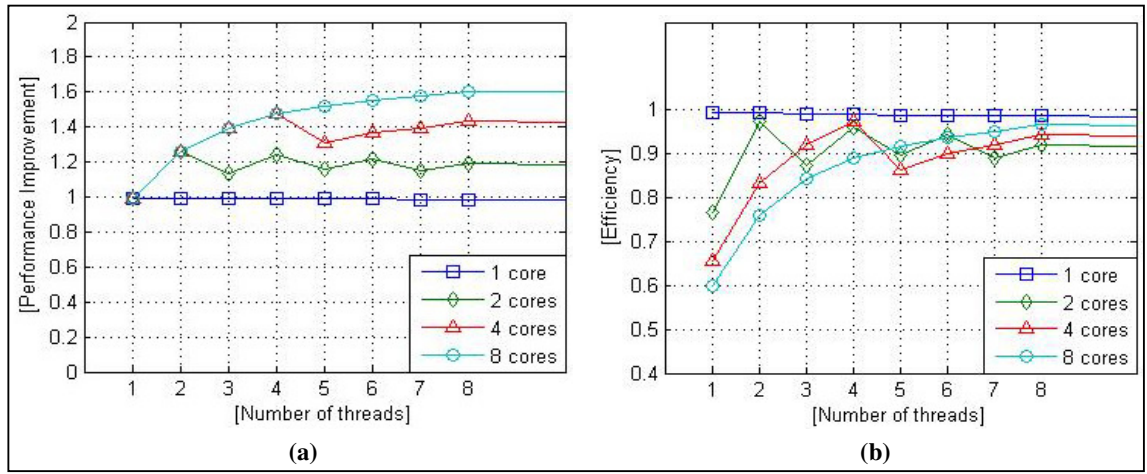

**Fig.8:** (a) Performance evaluation of Meisjter Algo. (b) Efficiency evaluation of Meisjter Algo.

The measurements were done on 2D binary image (512*512). If we can get a satisfactory outcome for this standard, it will be the same for smaller size images. View cache size limits, larger image will not be tested. Figure 7 shows that number of instructions to compute Euclidian distance drops from an average of $9.5 \times 10^8$ using 4SED algorithm down to $7.6 \times 10^8$ ms with Meijster algorithm. Despite the passage from a sequential version running on single core to a parallel version running on 8 processors, acceleration is only multiplied by 1.6 as shown in figure 8 (a). This can be explained by the choke point between columns processing and lines processing. Waiting time between these two treatments significantly penalizes acceleration. Figure 8 (b) shows that efficiency variation depends on the number of threads. It is also proportional to the number of processors. Moving to 3, 5 or 7 threads (odd number) decreases significantly the efficiency which reaches its maximum each time that the number of threads is equal the number of processors.

### 4.2 Thinning and thickening computing

Algorithms of thinning and thickening are almost the same. The only difference between them is the following: in thinning algorithm, destructible points are detected then their values are lowered. In thickening algorithm, constructible points, are detected then their values are increased. For parallelization, we will apply the same techniques introduced in [25]. We propose a similar version using two loops. Target points are initially detected then their value lowered or enhanced according to appropriate treatment. The set of their eight (or four) neighbors are copied into a "buffer" and rechecked. This treatment is repeated until stability. In the following, we present an adapted version of Couprie's thinning algorithm.

**Algorithm 7: Adapted Version Thinning Algo.**

1. **while** ($input[x]$ is destructible) **do**
2.     $push(x, stack1)$
3.     $x \leftarrow x+1$
4. **endwhile**
5. $output \leftarrow input$
6. **While** $(stack1 \neq \emptyset) \land (\max_{iter} > 0)$ **do**
7.     **While** $(stack1 \neq \emptyset)$ **do**
8.         $x \leftarrow pop(stack1)$
9.         **if** ($output[x]$ is destructible) **then**
10.             $output[x] \leftarrow reduce\_pt(x)$
11.             $push(x, stack2)$
12.         **endif**
13.     **end while**
14.     **While** $(stack2 \neq \emptyset)$ **do**

```
15.         x ← pop(stack2)
16.         v ← neighbors(x)
17.         i ← 0
18.         While (i < 8) do
19.             if (v[i] ∉ stack1) then
20.                 push(v[i], stack1)
21.             endif
22.         endwhile
23.     endwhile
24.     max_iter ← max_iter - 1
25. Endwhile
```

Unfortunately direct application of introduced parallel processing is not possible with the set of all points. Some points, called critical points, cannot be eliminated in parallel because initial topology of the image may be broken. Figure 9 illustrates this case: Critical points of an input image (a) are identified in (b). If these points are deleted in one iteration (c) topology necessary is broken (d).

To resolve this problem, we propose that research areas assigned to each thread must be composed of at least six lines (of the image). Each thread will use two buffers to treat each three lines thus four buffers are used to treat six lines as shown in 9 (e).

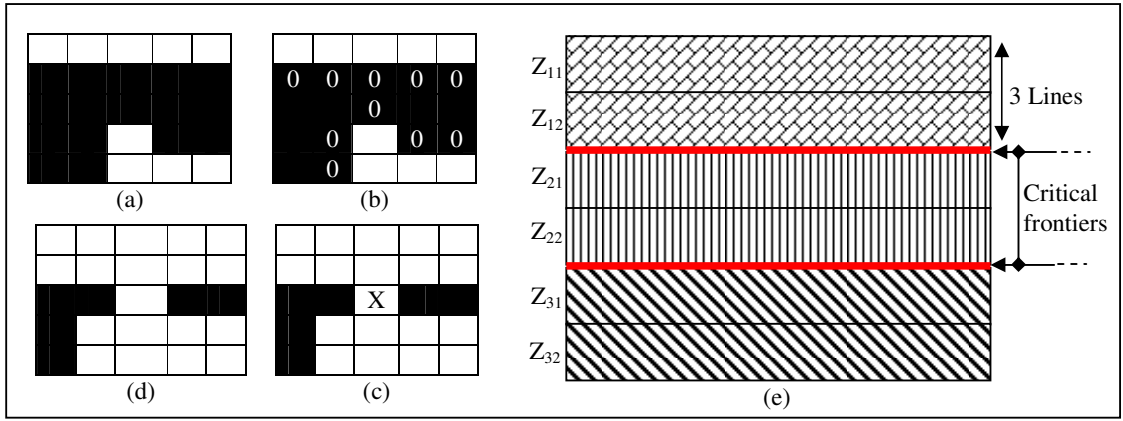

**Fig.9:** (a) (b) (c) (d) Critical point illustration (e) Research Area Assignment

Through this organization threads can start running in parallel on $Z_{11}$, $Z_{21}$ and $Z_{31}$. Once processing is completed threads can restart running on $Z_{12}$, $Z_{22}$ and $Z_{32}$. In some cases, a neighbor of a destructible point is detected on the border of a contiguous area. To prevent that such neighbor escape to recheck, it must be injected to buffer of the right thread. Let's suppose that a point $p \in Z_2$ is considered as destructible by T2, so its value will be lowered and its four neighbors $\{v_1, v_2, v_3, v_4\}$ should be rechecked. Neighbors $\{v_1, v_2, v_4\}$ belong to $Z_2$ so they will be push in T2 buffers. The neighbor $\{v_3\}$ belongs to $Z_3$ so it will stack T3 buffers.

Performance evaluation of introduced adapted version of Couprie's algorithm is shown in figure 10. On eight cores architecture, acceleration does not exceed 3.4. Such moderate result can be explained by critical borders processing. Regarding efficiency, the best performance is achieved when the number of thread is equal to the number of processors. If this equality is not ensured, the efficiency decreases. The problem threads' add number still persists.

The next step is to combine the parallel version of Meijster algorithm and the adapted version of Couprie's algorithm to build the parallel processing of topological smoothing.

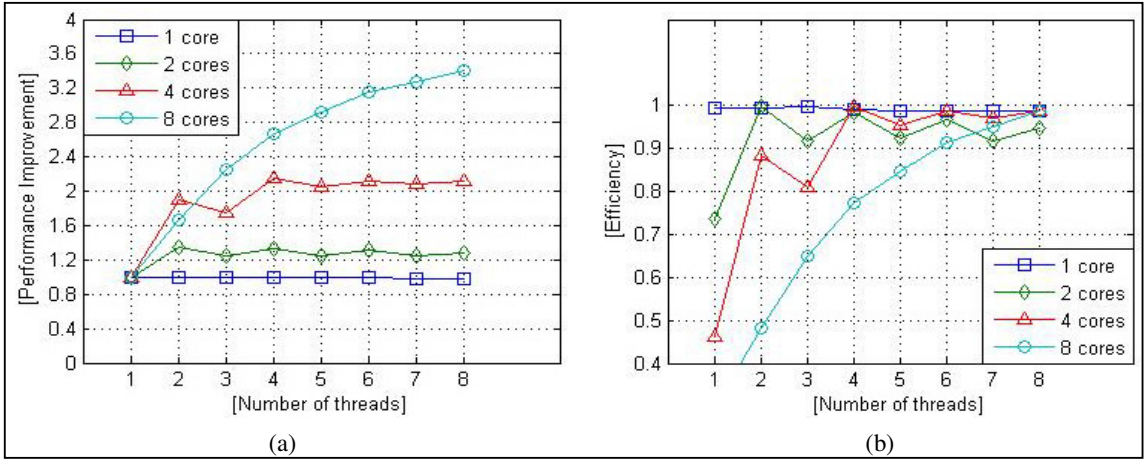

**Fig.10:** (a) Performance Evaluation of Couprie's Algorithm (b)Efficiency Evaluation of Couprie's Algorithm.

### 4.3 Global analyses

In this section, we present a global evaluation of the parallel smoothing operator. We start by presenting performance evaluations in terms of acceleration and efficiency. Then we evaluate cache memory consumption.

#### 4.3.1 Execution time

We implemented two versions of the proposed parallel topological smoothing algorithm, the first one using 'Symmetric Multiprocessing' scheduler and the second one using 'basic-NPS' scheduler. Wall-clock execution times for numbers of threads equal to 1, 2, 4, 8, and 16 were determined. The minimum value of 2 timings was taken as most indicative of algorithm speed. The measurements were done on 2D binary image (512*512). Results of the second implementation on the eight-core are shown in the following figure.

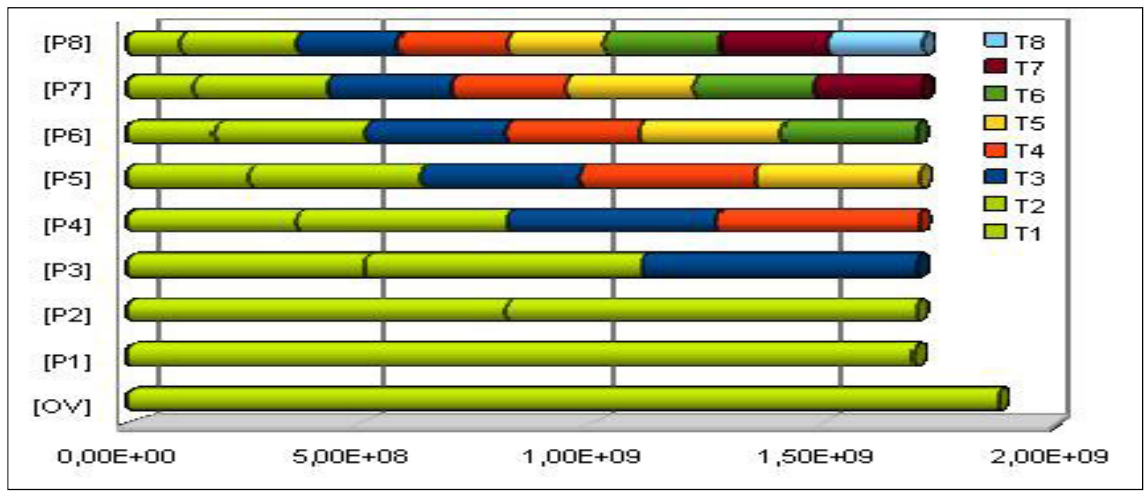

**Fig. 11:** Number of Instructions and tasks distribution using 'Basic-NPS'

We note that number of instructions drops from an average of 1879 $\times 10^8$ FI with a single thread down to 1652 $\times 10^8$ ms with 8 threads. As expected, the speed-up for the second implementation using 'basic-NPS' scheduler is higher than for the one using "Symmetric Multiprocessing" scheduler, thanks to balanced distribution of tasks. A remarkable result about speedup is also shown in figure 12 (a). In fact, speed-up increases as we increase the number of threads beyond the number of processors in our machine (eight cores). In the first implementation, using "Symmetric Multiprocessing" scheduler, the speedup at 8 threads is 1.9 ± 0.01. However, for the second implementation, using our scheduler, the speedup has increased to 5.2 ± 0.01. Another common result between different architecture is stability of execution time on each n-core machine since the code uses n or more threads.

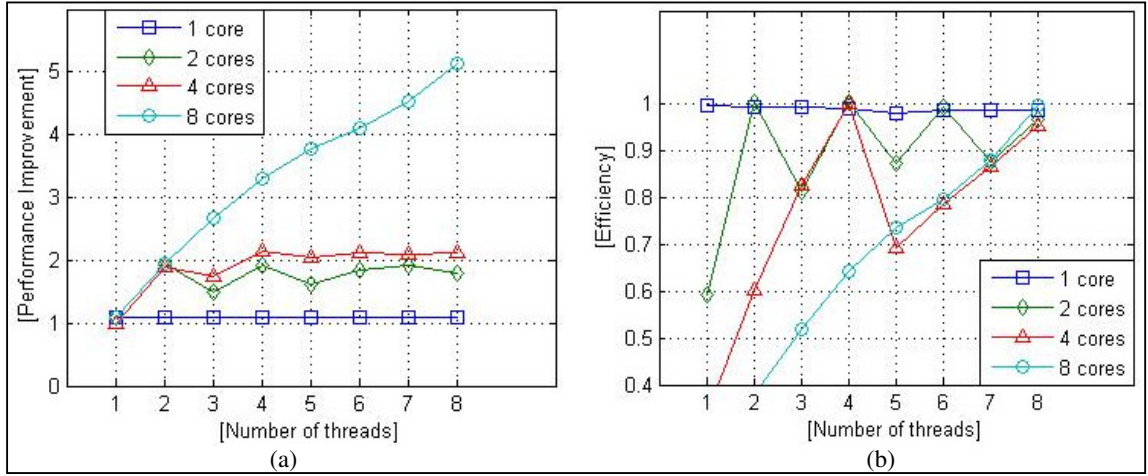

**Fig. 12: (a)** Global Performance improvement **(b)** Global efficiency improvement

For better readability of our results, we tested also efficiency of our algorithm on various architectures (see figure 12(b)) using the $\psi(n)$ formula introduced earlier. For parallel time ratio we used best obtained time with 8 threads ('basic-NPS' scheduler).

### 4.3.2 Cache Memory Evaluation

As memory access is a principal bottleneck in current-day computer architectures, a key enabler for high performance is masking the memory overhead. If we starts from basic theory that two classic cache design parameters dramatically influence the cache performance: the block size and the cache associativity. So the simplest way to reduce the miss rate is to increase the block size even it increases the miss penalty. The second solution is to decrease associatively in order to decrease hit time thus to retrieve a block in an associative cache, the block must be searched inside of an entire set since there is more than one place where the block can be stored.

Unfortunately, we are dealing with non-reconfigurable architectures with caches whose associativity and block size are predefined by the manufacturer. Nowadays, new approaches to reduce cache miss are developed such as taking advantage of locality of references to memory or using aggressive multithreading so that whenever a thread is stalled, waiting for data, the system can efficiently switch to execute another thread. Despite their power, the application of both approaches remains limited. In fact, applications of locality approach still experimental even with Larrabee technology introduced by Intel. And the aggressive multithreading approach has been specially designed for graphics processing engines, which manage thousands of in-flight threads concurrently. So it is not recommended for general SMP machines with limited number of processors and threads. With all these limitations, the most intuitive solution is to rely on the scheduling. Thanks to our basic-NPS scheduler, we have balanced the charges then prevent context switching thus we minimize caches misses.

|  |  | Intel P4 | Intel Dual C. T1400 | Intel C2 Quad Q9550 | Intel Xeon E5405 |
|---|---|---|---|---|---|
| Num. of processor |  | 1 | 2 | 4 | 2 x 4 |
| SMT |  | Yes | Yes | Yes | Yes |
| Frequency |  | 3,4 GHz | 1,73 GHz | 2,83 GHz | 2,00 GHz |
| L1 Instr. Cache | Size | 16Kb | 32Ko | 32Ko | 32Ko |
|  | Asso. | 8-way | 8-way | 8-way | 8-way |
|  | Block size | 32byte | 32byte | 32byte | 32byte |
| L1 Data Cache | Size | 16Kb | 32Ko | 32Ko | 32Ko |
|  | Asso. | 8-way | 8-way | 8-way | 8-way |
|  | Block size | 64byte | 64byte | 64byte | 64byte |
| L2 Cache | Size | 2Mb | 512Kb | 6Mb | 6Mb |
|  | Asso. | 8-way | 8-way | 8-way | 8-way |
|  | Block size | 64byte | 64byte | 64byte | 64byte |
| RAM size |  | 1Gb | 2Gb | 2Gb | 8Gb |

**Tab.2:** Hardware configuration

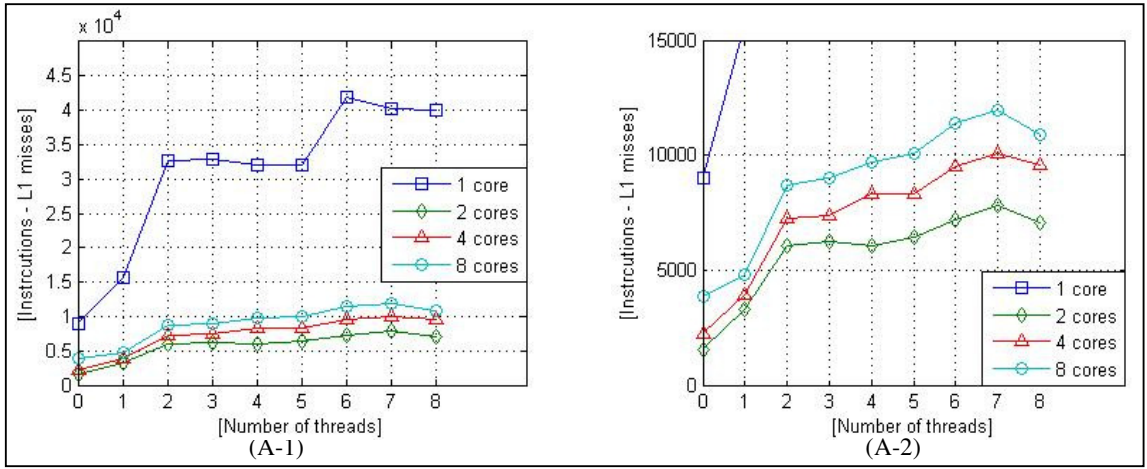

**Fig. 13:** (**A**) Instruction - L1 misses

In the following we present our experimental analysis. We consider a commonly used Intel processor configuration (More details are given by table 2). Number of processor varies from one to eight. The frequency varies between 1,73 GHz and 3,4 GHz. The L1 caches have at least a 32-byte block size, while capacity vary between 16 Kbytes and 32 Kbytes, and for the associativity, only eight ways is considered. The L2 caches have at least a 64-byte block size, while capacities vary between 512 Kbytes and 6 Mbytes, and the associativity varies between two and twenty four ways.

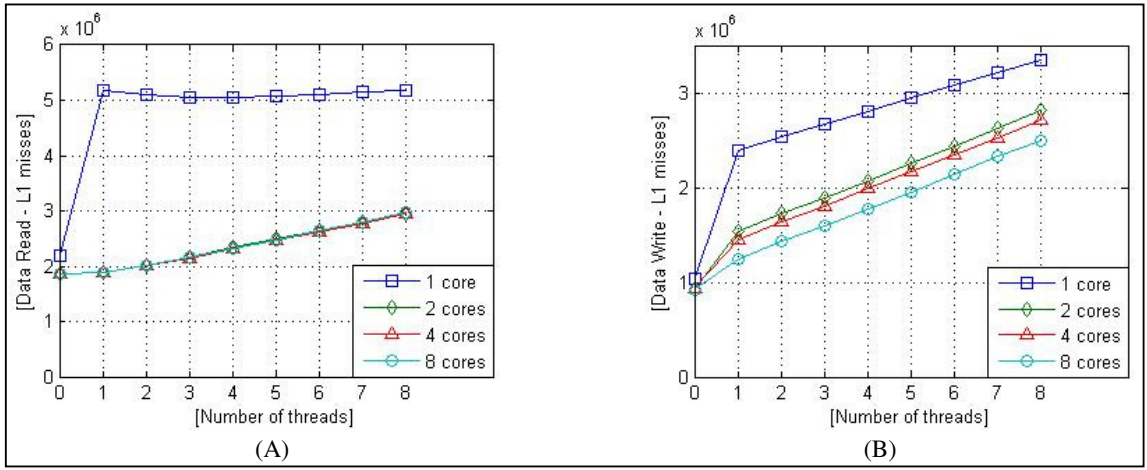

**Fig. 14:** (**A**) Data Read - L1 misses (**B**) Data Write - L1 misses

The scheduler relies on our basic-NPS scheduling policy. As a result of this experiment, see figure 13 (A-1), we found that three performance regions are clearly evident: In the leftmost region, as long as the cache capacity can effectively serve the growing number of threads, increasing the number of threads improves performance, as more processors are utilized. This area is generally identified as cache-efficiency zone. At some point, the cache becomes too small for the growing stream of access requests, so memory latency is no longer masked by the cache and instruction cache misses reduce more moderately. As the number of available threads again increases, the multithread efficiency zone (on the right) is reached, where adding more threads improves performance up to the maximal performance of the machine, or up to the bandwidth wall. Balanced workloads offer higher locality and better exploit the cache and hence expand the cache efficiency zone to the right and up. An outstanding example is given by the following table which summarizes number of L1 instruction misses on Intel Dual Core T1400 architecture using SMP scheduling policy and Basic-NPS scheduling policy. We note that number of instruction misses drops from an average of 18844 L1 Instr. misses (using SMP) with two threads down to 6030 L1 Instr. misses (using Basic-NPS) usually with two threads. Here success rate is largely above the average of 50%. The same rate will be practically maintained when increasing the number of threads.

| Number of threads |                      | 1     | 2     | 3     | 4     | 5     | 6     | 7     | 8     |
|-------------------|----------------------|-------|-------|-------|-------|-------|-------|-------|-------|
| Instr.            | Sym. Multi. Scheduler | 10298 | 18844 | 19476 | 18638 | 19726 | 20058 | 20324 | 18946 |
| L1 misses         | Basic-NPS scheduler  | 3307  | 6030  | 6262  | 6035  | 6437  | 7202  | 7804  | 7085  |

**Tab.3:** L2 – Instructions Misses (Symmetric Multiprocessing scheduler vs. Basic-NPS scheduler)

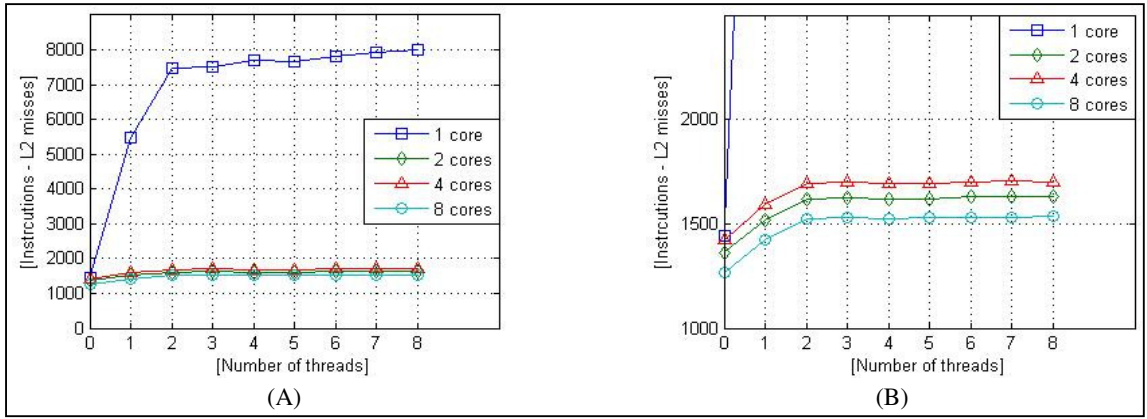

(A)  (B)

**Fig. 15:** Instruction – L2 misses (B): zoom on (A)

Moreover, the shape of the performance curve depends on how fast the cache hit rate degrades as a function of the number of threads. Any success access to L1 will eliminate an attempt to access to L2 thus performance curve, Fig. 15, will evaluate in the same way. By reducing the number of cache miss from instruction cache, processor or thread of execution has not to wait (stall) until the instruction is fetched from main memory which immediately impact execution time.

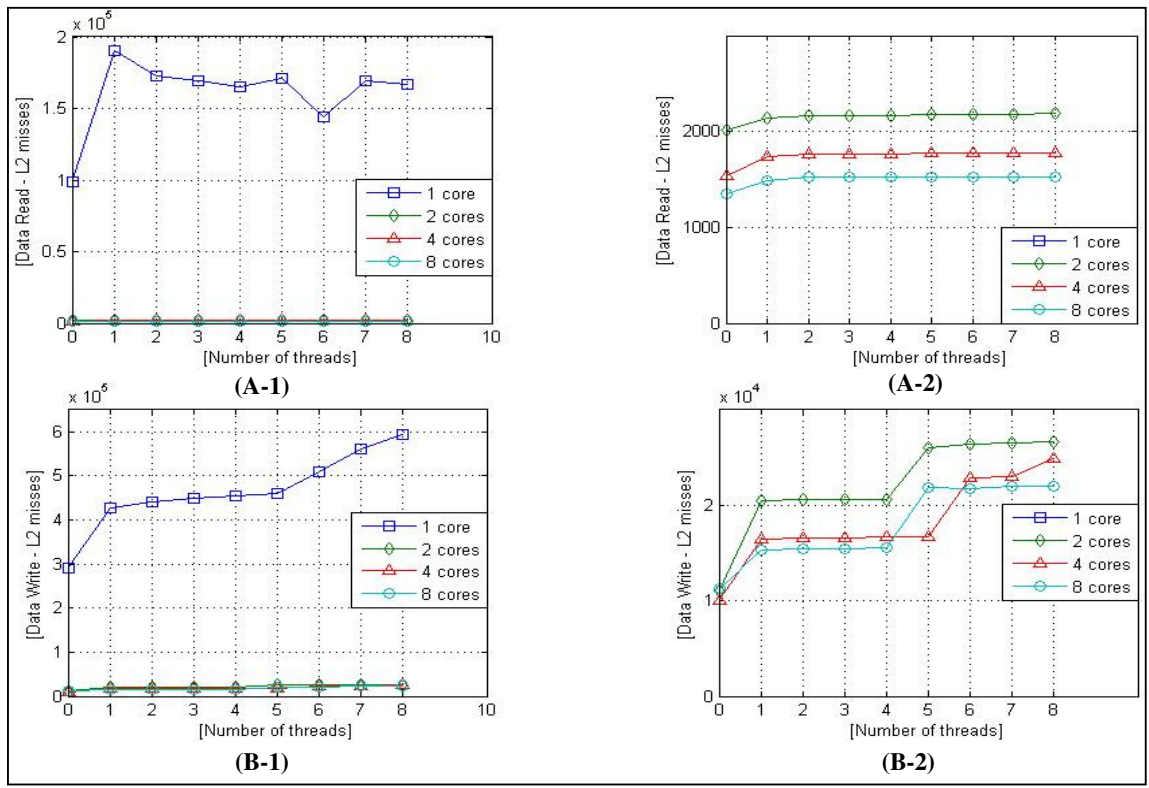

**(A-1)**  **(A-2)**  **(B-1)**  **(B-2)**

**Fig. 16:** (**A**) Data Read – L2 misses (**B**) Data Write – L2 misses

Figures 14 (A) and 16 (A-1) show so much load balancing and implicitly context switching between processes can affect performance in terms of reading data from caches. However, improvement in writing data, see Fig. 14 (B) and Fig. 16 (B-1), in two caches remains modest. When there are more computation instructions per memory access, performance climbs more steeply with additional threads. This is because as more instructions are available for each memory access, fewer threads are needed to fill the stall time resulting from waiting for memory.

## 5 Conclusion

Topological characteristics are fundamental attributes of an object. In many applications, it is mandatory to preserve or control the topology of an image. Nevertheless, the design of transformations which preserve both topological and geometrical features of images is not an obvious task, especially for parallel processing.

In this paper, we have presented a new parallel computation method for topological smoothing through combining parallel computation of Euclidean Distance Transform using Meijster algorithm and parallel Thinning–Thickening processes using an adapted version of Couprie's algorithm.

We have also presented a new parallelization strategy called SDM (Split Distribute and Merge). Proposed strategy is partially based on divide and conquers principle associated to event-based coordination techniques. Further than smoothing operator, SDM Strategy can be applied for a large class of topological operators as we shown in section 3.1. In addition to identified conditions during splitting step, we introduced an adapted scheduler called basic-NPS (Basic - Non Preemptive Scheduler) able to distribute in balanced way a set of active tasks on available processors. Finally we introduced an adapted merging policy designed especially for dynamic system evolving until stability.

Parallel topological operator computation poses many challenges, ranging from parallelization strategies to implementation techniques. We tackle these challenges using successive refinement, starting with highly local operators, which process only by characterizing points and then deleting target pixels, and gradually moving to more complex topological operators with non-local behavior. In future work, we will study parallel computation of the topological watershed [14].